\documentclass[aps,prb,preprintnumbers,twocolumn,showpacs]{revtex4}

\usepackage{psfrag,graphicx}
\usepackage{dcolumn}
\usepackage{amsmath,amssymb}
\usepackage{bm}
\usepackage{amsfonts,amssymb,amsmath}        % for math symbols.

\newcommand{\be}{\begin{equation}}
\newcommand{\ee}{\end{equation}}
\newcommand{\bq}{\begin{eqnarray}}
\newcommand{\eq}{\end{eqnarray}}

\newcommand{\Alpha}{\mbox{\boldmath${\alpha}$}}
\newcommand{\SIGMA}{\mbox{\boldmath${\sigma}$}}
\newcommand{\ket}[1]{\left | \, #1 \right\rangle}

\newcommand{\rf}[1]{(\ref{#1})}
\begin{document}

\title{Topological phase transitions driven by gauge fields in an exactly solvable model}

\author{Ville Lahtinen and Jiannis K. Pachos}
\affiliation{School of Physics and Astronomy, University of Leeds, Woodhouse Lane, Leeds LS2 9JT, UK}
\date{\today}

 \begin{abstract}

We demonstrate the existence of a new topologically ordered phase in Kitaev's honeycomb lattice model. This new phase appears due to the presence of a vortex lattice and it supports chiral Abelian anyons. We characterize the phase by its low-energy behavior that is described by a distinct number of Dirac fermions. We identify two physically distinct types of topological phase transitions and obtain analytically the critical behavior of the extended phase space. The Fermi surface evolution associated with the transitions is shown to be due to the Dirac fermions coupling to chiral gauge fields. Finally, we describe how the new phase can be understood in terms of interactions between the anyonic vortices.

\end{abstract}


\maketitle

\section{Introduction}

Unlike other phases of matter, topologically ordered phases of many-body quantum systems can not be characterized by local symmetries. Instead, they are described by long-range properties, which are responsible for the emergence of anyons, exotic quasiparticles with fractional statistics. Such phases occur in the celebrated fractional Quantum Hall systems \cite{Nayak} or in a variety of lattice models \cite{Brennen}. The latter offer microscopic control and flexibility to study topologically ordered phases. Among them a pioneering role has been played by Kitaev's honeycomb lattice model \cite{Kitaev05}, where the vortices can behave as non-Abelian anyons. The analytic tractability of the model has enabled the explicit demonstration of their various characteristics including fusion rules \cite{Lahtinen}, braid statistics \cite{Lahtinen09}, topological degeneracy \cite{Kells09}, edge states \cite{Lee} and entanglement entropy \cite{Yao10}. Several experimental realizations of the model have also been proposed \cite{Micheli05}.

Here we demonstrate the existence of a new topological phase in the honeycomb lattice model, which has been anticipated by Kitaev \cite{Kitaev06}. It appears for a fully packed lattice of vortices, when they behave as non-Abelian anyons. The ground state of this configuration is characterized by Chern number $\nu=\pm2$, which implies the emergence of a new phase that supports chiral Abelian anyons. To understand the transitions between the different topological phases, we study the evolution of the Fermi surface, which describes the long-range properties of the model. By considering the low-energy field theory of Dirac fermions, we show how two distinct types of topological phase transitions, i.e. different changes in the Fermi surface topology \cite{Volovik}, occur due to coupling to chiral gauge fields. We identify these to be due to dimerization and staggering of the model's couplings, both which can be realized in the proposed optical lattice experiments \cite{Micheli05}. Finally, we derive analytically the phase boundaries for the new phase and illustrate how the transition between the non-Abelian and the chiral Abelian phase can be understood in the context of anyon-anyon interaction driven phase transitions \cite{Trebst}. Due to an equivalence between the models \cite{Chen08}, our results apply directly also to $p$-wave superconductors.

\begin{figure}[pt]
\includegraphics*[width=8cm]{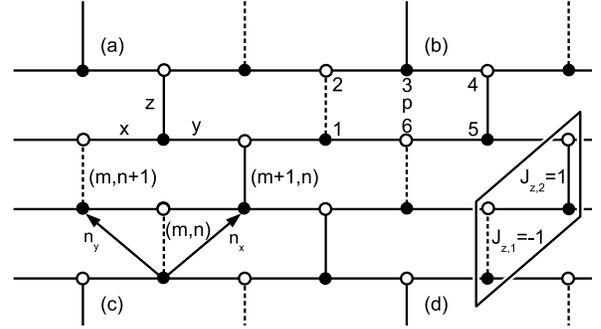}
\caption{\label{honeycomb} A brickwall representation of the bi-colorable honeycomb lattice. (a) The links are labeled $x$, $y$ or $z$ depending on their orientation. (b) Numbering of sites on plaquette $p$. (c) The elementary cell comprises of a black and a white site connected by a $z$-link. The cells are labeled by $\mathbf{k}=(m,n)$ in the orthonormal basis $\mathbf{n}_x = (1,0)$ and $\mathbf{n}_y = (0,1)$. (d) The full-vortex configuration is created by setting $J_{ij}=-1$ on alternating $z$-links in $x$-direction (dashed links) and $J_{ij}=1$ on all other links (solid links). The parallelogram shows the four sites belonging to the unit cell.}
\end{figure}

\section{The honeycomb lattice model}

\subsection{Mapping to free Majorana fermions}

Kitaev's model \cite{Kitaev05}, comprises of spin-$1/2$ particles residing
on the vertices of a honeycomb lattice. The spins interact according to the
Hamiltonian
\begin{eqnarray} \label{H}
    H = - \sum_{( i,j)} J_{ij} \sigma^\alpha_i \sigma^\alpha_j -
     K \sum_{( i,j,k)} \sigma^x_i \sigma^y_k \sigma^z_j,
\end{eqnarray}
where $J_{ij}=(J_\alpha)_{ij}$ are nearest neighbour couplings and $\alpha$ is either
$x$, $y$ or $z$ when $(ij)$ is an $x$, $y$ or $z$-link, respectively (see Figure
\ref{honeycomb}(a)). The second term is an effective magnetic field
of magnitude $K$, which explicitly breaks time-reversal invariance. The sum in this term runs over all next to nearest
neighbour triplets \cite{Lahtinen}.

When the spin operators are represented as
$\sigma^{\alpha}_i = i b^\alpha_i c_i$, where $c_i, b^x_i, b^y_i$ and
$b^z_i$ are Majorana fermions, the Hamiltonian takes the quadratic form
$H = \frac{i}{4} \sum_{i,j} \hat{h}_{ij} c_{i} c_{j}$, where
\begin{eqnarray} \label{h}
    \hat{h}_{ij} = 2J_{ij} \hat{u}_{ij} + 2 K \sum_{k}\hat{u}_{ik}
    \hat{u}_{jk}, \quad \hat{u}_{ij} = ib^\alpha_i b^\alpha_j.
\end{eqnarray}
The eigenstates $\ket{\Psi}$ of the original Hamiltonian \rf{H} are subject to the constraint
\be \label{D}
D_i \ket{\Psi} = \ket{\Psi}, \qquad D_i = b^x_i b^y_i b^z_i c_i.
\ee
Since  $[H, \hat{u}_{ij}] = 0$, the Hilbert space splits into sectors each
labelled by $u$, a certain pattern of eigenvalues $u_{ij}=\pm1$. However, as $D_i \hat{u}_{ij}=-\hat{u}_{ij} D_i$, $\hat{u}_{ij}$ should be understood as a classical $Z_2$ gauge field with
local gauge transformations $D_i$. The physical gauge-invariant sectors are labeled by the eigenvalues of the plaquette operators $\hat{w}_p = \hat{u}_{12}\hat{u}_{32}\hat{u}_{34}\hat{u}_{54}\hat{u}_{56}\hat{u}_{16}$, that satisfy $[H,\hat{w}_p]=0$ (see Figure \ref{honeycomb}(b)). These can be identified with gauge
invariant Wilson loop operators and their eigenvalues $w_p = - 1$ can be interpreted as having a {\em vortex} on plaquette $p$. We refer to the eigenvalue patterns $u = \{ u_{ij} \}$ and $w = \{ w_p \}$ as gauge and vortex sectors, respectively.

\subsection{Gauge / coupling configuration equivalence}

To study the model as the vortex sector is varied, it is convenient to absorb the gauge $u$ into the couplings. As can be seen from \rf{h}, the gauge choice $u_{ij}=\pm1$ can be locally regarded as the sign of the physical coupling $J_{ij}$. Therefore, the system can equivalently be viewed as being initially prepared in the vortex-free sector ($w_p=1$ on all plaquettes) that contains the ground state \cite{Lieb}, with the other vortex sectors emerging by tuning the signs of the coupling configurations $J = \{ J_{ij} \}$. This view is also physically motivated as it relates the creation and transport of the vortices to the local manipulation of the couplings \cite{Lahtinen09}.

Strictly speaking one should also reverse the sign of the effecticve magnetic field couplings $K$ locally. However, when $K$ approximates an external magnetic field, i.e. when $K \ll |J_\alpha|$, considering only the signs of $J_\alpha$ is a good approximation. Furthermore, we will see later that a non-zero $K$ is required only to open a spectral gap. It does not affect the Fermi surface topology that characterizes the different phases.

\subsection{The full-vortex sector}

When the couplings satisfy $J_x \approx J_y \approx J_z$ and $K >0$ and there are only a few vortices in the system, the vortices behave as strongly interacting non-Abelian Ising anyons \cite{Lahtinen}. The chiral Abelian phase emerges in this same coupling regime when one places a vortex on every plaquette, i.e. considers the a full-vortex sector ($w_p=-1$ on all plaquettes). It can be created, for instance, by using a gauge where $u_{ij}=\pm 1$ alternates on all $z$-links in direction $\mathbf{n}_x$, while $u_{ij}=1$ on all other links. The lattice geometry and the corresponding unit cell containing four sites are illustrated in Figures \ref{honeycomb}(c)-(d). Normalizing $|J_\alpha|=1$, the gauge above is then equivalent to setting inside the unit cell $ J_x=1, J_y=1$ and $(J_{z,1}, J_{z,2})=(-1,1)$, which corresponds to staggered couplings on $z$-links in direction $\mathbf{n}_x$. The system is translationally invariant with respect to $(2\mathbf{n}_x,\mathbf{n}_y)$ and a Fourier transformation gives the Hamiltonian $H = \int_{BZ} \textrm{d}^2 \mathbf{p} \ \mathbf{c}_{\mathbf{p}}^\dagger H_\mathbf{p} \mathbf{c}_{\mathbf{p}}$, where
\begin{eqnarray} \label{H1}
    H_\mathbf{p} = \left( \begin{array}{cc} h_{bb} & h_{bw} \\ h^\dagger_{bw} & -h^\textrm{T}_{bb} \end{array}\right), 
\end{eqnarray}
with
\begin{eqnarray}
    h_{bw} = \left( \begin{array}{cc} i ( e^{ip_x} + e^{ip_y}) & i \\ i & i( -e^{ip_x} + e^{ip_y}) \end{array}  \right)
\end{eqnarray}
and
\begin{eqnarray}
    h_{bb} = K \left( \begin{array}{cc} \sin(p_x-p_y) & \sin(p_y)-i\cos(p_x) \\ \sin(p_y)+i\cos(p_x) & -\sin(p_x-p_y) \end{array}  \right).
\end{eqnarray}
The integral is over the first Brillouin zone ($-\frac{\pi}{2} \leq p_x \leq \frac{\pi}{2}$, $-\pi \leq p_y \leq \pi$) and  $\mathbf{c}_{\mathbf{p}}=(c_{b,\mathbf{p}}, c_{b,\mathbf{p}+\pi\mathbf{n}_x}, c_{w,\mathbf{p}}, c_{w,\mathbf{p}+\pi\mathbf{n}_x})$. The indices $(b,w)$ and ($\mathbf{p},\mathbf{p}+\pi\mathbf{n}_x$) refer to the two triangular sublattices of the honeycomb lattice and the two $z$-links inside the unit cell, respectively.

This Hamiltonian can be readily diagonalized, with the lengthy expressions for the eigenvalues $\pm E_{1,\mathbf{p}}$ and $\pm E_{2,\mathbf{p}}$ given in \cite{Lahtinen}. In contrast to the vortex-free sector that has two Fermi points \cite{Kitaev05}, the low-energy behavior when $K=0$ is now dominated by the four Fermi points ($E_{1,\mathbf{Q}}=0$) located at $\mathbf{Q}^1_\pm = \mp (\frac{\pi}{3},\frac{\pi}{6})$ and $\mathbf{Q}^2_\pm = \pm (-\frac{\pi}{3}, \frac{5\pi}{6})$. Their number and locations follow from the symmetries of \rf{H1},
\begin{eqnarray}
	\Gamma_1 = \sigma^y \otimes \openone: \qquad \Gamma_1 H_\mathbf{p} \Gamma_1^\dagger & = & H_{-\mathbf{p}}, \label{SLS1} \\
	\Gamma_2 = \sigma^z \otimes \sigma^y: \qquad \Gamma_2 H_\mathbf{p} \Gamma_2^\dagger & = & H_{\mathbf{p}+\pi\mathbf{n}_y} \label{SLS2},
\end{eqnarray}
which hold also for non-zero $K$. $\Gamma_1$ acts only on the sublattice indices ($b,w$) and descibes a symmetry of the honeycomb lattice geometry. It is responsible for the doubling of Fermi points in the vortex-free sector. The new symmetry $\Gamma_2$ acts also on the indices ($\mathbf{p},\mathbf{p}+\pi\mathbf{n}_x$) that correspond to the two $z$-links inside the unit cell. Exchanging these links maps $(J_{z,1}, J_{z,2})=(-1,1) \to (1,-1)$, which preserves the full-vortex sector. Although $\Gamma_2$ is a lattice symmetry, it can be equivalently viewed as an emergent global $Z_2$ gauge symmetry as the corresponding gauges are inequivalent under the local gauge transformations $D_i$. This new symmetry is responsible for the further doubling of the Fermi points in the full-vortex sector. It changes the Fermi surface topology and thus implies a new phase.

\section{Fermi surface analysis}

\subsection{The low-energy theory of Dirac fermions}

To characterize this new phase and study its emergence from the phase with non-Abelian anyons, we consider its low-energy theory. The Hamiltonian \rf{H1} is expanded around each Fermi point $\mathbf{Q}^i_\pm$ by writing $\mathbf{p}=\mathbf{Q}+\mathbf{k}$, with $|\mathbf{k}| \ll 1$. In general one obtains 
\be
	H_{\mathbf{Q}} = H^0_{\mathbf{Q}}+H^x_{\mathbf{Q}}k_x + H^y_{\mathbf{Q}}k_y + \mathcal{O}(k^2),
\ee
for some $4 \times 4$ matrices $H^\eta_{\mathbf{Q}}$. A projection onto the 2-dimensional low-energy subspace is given by
\be
 \bar{H}_{\mathbf{Q}} = PU_{\mathbf{Q}} H_{\mathbf{Q}} U_{\mathbf{Q}}^\dagger P,
\ee
where $U_\mathbf{Q}$ diagonalizes $H^0_{\mathbf{Q}}$ and $P$ projects onto the zero eigenvalue states of $U_\mathbf{Q}H^0_{\mathbf{Q}}U_\mathbf{Q}^\dagger$. Keeping only the first order terms in $\mathbf{k}$ and projecting onto the low-energy states, we obtain
\begin{eqnarray} \label{H2}
 \bar{H}_{\mathbf{Q}^i_\pm} \approx \SIGMA_\pm^i \cdot \mathbf{k}^i \mp \sigma^z \frac{K}{2\sqrt{3}},
\end{eqnarray}
where $\SIGMA_\pm^i = (\sigma^x, \pm (-1)^i \sigma^y)$, $\mathbf{k}^i = ( \frac{a_i k_x-k_y}{1+a_i}, \frac{k_x-a_i k_y}{1+a_i})$ and $a_i=2-(-1)^{i}\sqrt{3}$. This Hamiltonian describes massive Dirac fermions in two spatial dimensions. We interpret the mass being due to the scalar field $K$, which couples chirally, i.e. with a different sign, at the different Fermi points. 

\subsection{Fermi surface topology and the Chern number}

 \begin{figure}[t]
 \includegraphics*[width=7cm]{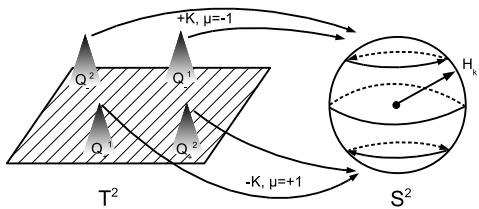} 
\caption{\label{winding} The normalized Hamiltonian \rf{H1} defines a mapping from the Brillouin
zone, which is topologically a torus $T_2$, to a unit sphere $S_2$. As $K$ contributes to the low-energy theory only around the Fermi points, the map is reduced to the contributions \rf{H2} from each. Depending on the sign of
the term $\pm K \sigma^z$, the neighbourhoods of $\mathbf{Q}_\pm^i$ are mapped to either lower or upper hemisphere.
As the ones with same orientations $\mu$ end up to same hemispheres, the map winds
around the unit sphere twice when viewed from the origin.}
\end{figure}

The chiral coupling of $K$ implies a chiral phase, which can be characterized by non-zero Chern number \cite{Haldane, Read00}. We can relate the low-energy theory to the Chern number as follows. As the contribution of $K$ vanishes away from the Fermi points, the four Hamiltonians \rf{H2} define an orientation preserving mapping from a torus (the first Brillouin zone) to a surface enclosing the origin (coordinates given in basis $\{ \sigma^\alpha \}$). The degree of this map gives the Chern number \cite{Kitaev05,Hsiang}. The orientation of Fermi point $\mathbf{Q}$ can be characterized by a winding number \cite{Wen89}
\be
 \mu_Q = \frac{1}{4\pi i} \oint_{C_Q} \textrm{Tr}\left( \Gamma H_{\mathbf{p}}^{-1}\textrm{d}H_{\mathbf{p}} \right),
\ee
where $C_Q$ is a loop in the momentum space around Fermi point $\mathbf{Q}$ and $\Gamma=\sigma^z \otimes \openone$.
Due to the chiral coupling of $K$, the neighborhoods of both $\mathbf{Q}^i_+$ ($\mathbf{Q}^i_-$) with orientations $\mu_{Q_+^i}=+1$ ($\mu_{Q_-^i}=-1$) are mapped to the lower (upper) hemisphere, with rest of the Brillouin zone being mapped to the equator. This is illustrated in Figure \ref{winding}. For four Fermi points the map winds twice around the origin giving the Chern number $\nu=-2$. We have verified this by calculating $\nu$ also using the exact eigenstates \cite{Fukui}, as well as observed two non-degenerate edge states when a system of $10^3$ lattice sites is placed on a cylinder \cite{Hatsugai}. The Chern number $\nu=-2$ implies that the vortices in this phase are certain chiral Abelian anyons as cataloged for free fermion systems by Kitaev \cite{Kitaev05}.

 \begin{figure}[t]
 \begin{tabular}{cc}
 \includegraphics*[width=4cm]{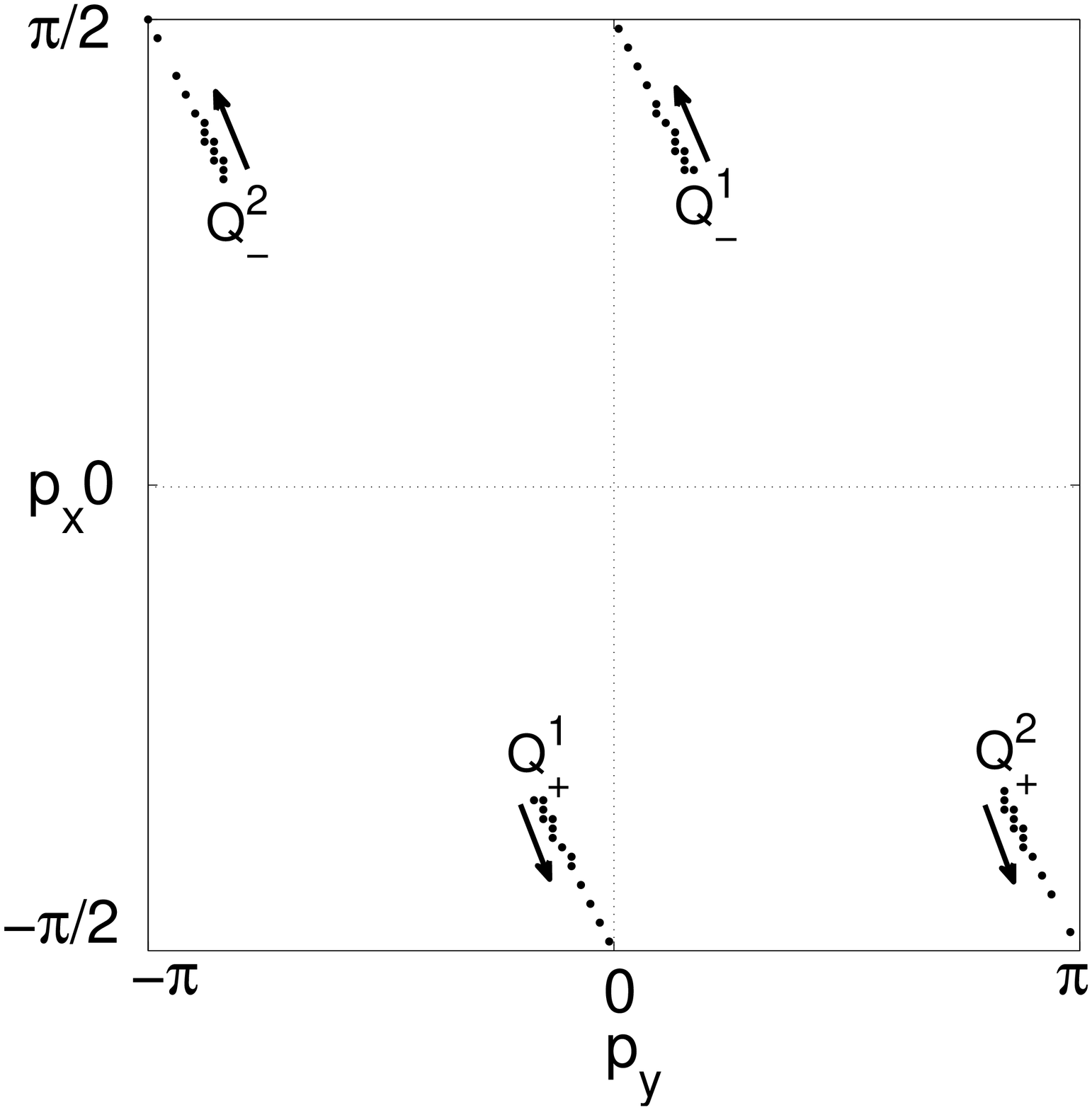} & \includegraphics*[width=4cm]{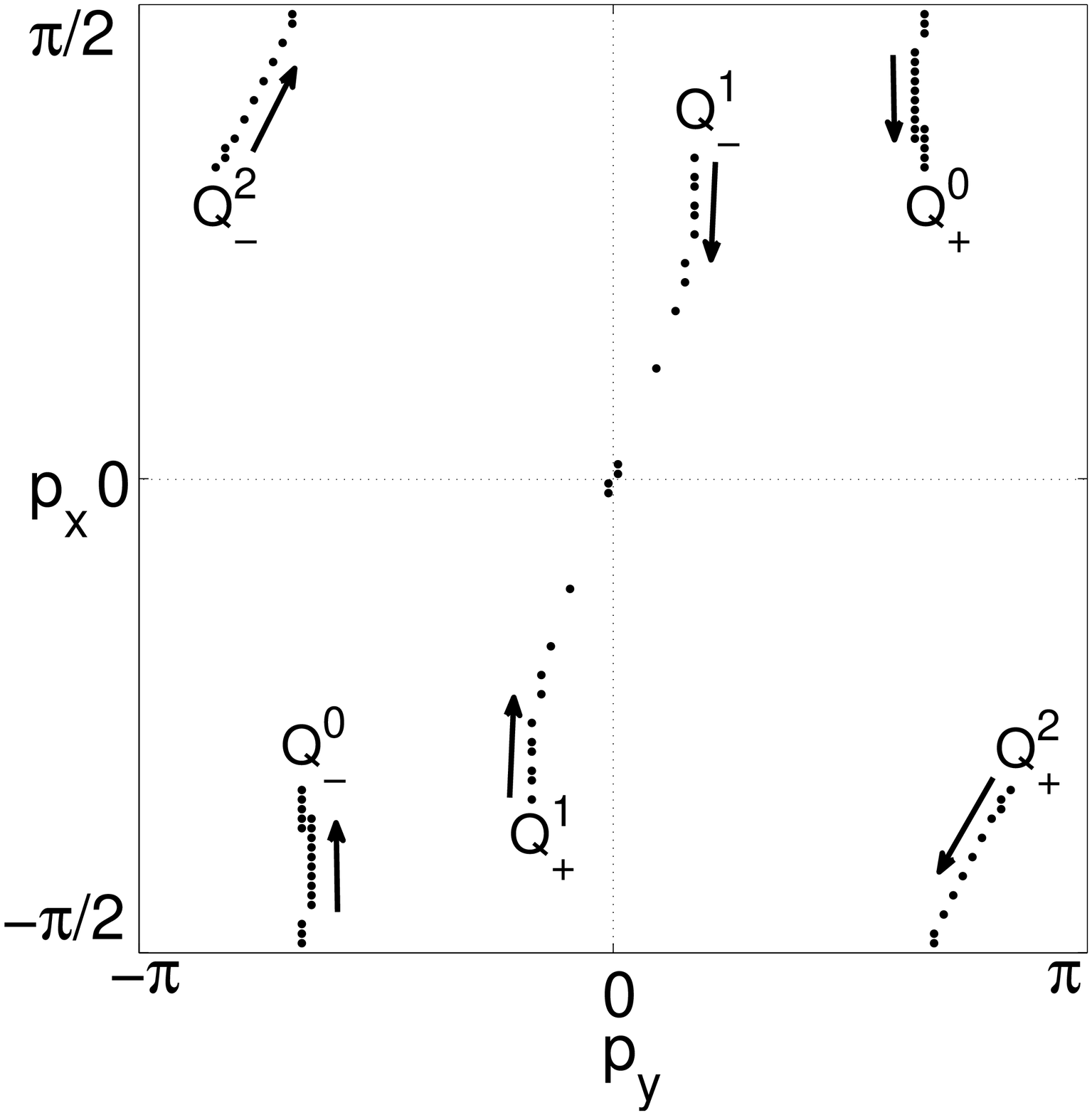} \\
(a) & (b)
\end{tabular}
\caption{\label{FP} A numerical study of the evolution of Fermi point locations (black dots) across the first Brillouin zone when (a) $\delta J_1 = 0 \to \infty$ ($\nu=-2 \to 0 $ transition due to dimerization) and (b) $\delta J_2 =0 \to 1$ ($\nu=-2 \to -1$ transition due to removal of coupling staggering). The Fermi points move in the directions of the arrows and the annihilations occur when $\delta J_1^c = \sqrt{2}-1$ and $\delta J_2^c = \frac{1}{4}$, respectively.}
\end{figure}

\subsection{Emergent gauge fields and Fermi point transport}

The phase boundaries of this new chiral Abelian phase can be obtained by studying the effect of tuning the couplings $J$ on the low-energy field theory. This allows one to obtain analytically the critical parameter values when the Fermi surface topology changes, i.e. when Fermi points are either created or annihilated. To this end we consider the system initially in the $\nu=-2$ phase and set $K=0$. The phase with non-chiral Abelian toric code anyons ($\nu=0$) appears due to dimerization when, for instance, $|J_z| \gg |J_x|, |J_y|$ in \rf{H} \cite{Kitaev05}. This can be modelled in the full-vortex Hamiltonian \rf{H1} as the perturbation 
\be \label{dH1}
	\delta H_1=-\delta J_1 \sigma^y \otimes \sigma^x. 
\ee
Projecting this into the low-energy subspace and combining the linearized Hamiltonians \rf{H2} for the paired Fermi points as $\bar{H}_{\mathbf{Q}^i}=\text{diag}(\bar{H}_{\mathbf{Q}^i_+},\bar{H}_{\mathbf{Q}^i_-})$, we obtain for small perturbations ($\delta J_1 \ll 1$)
\begin{eqnarray} \label{H4}
\bar{H}_{\mathbf{Q}^i} + \delta \bar{H}_{j} =  \Alpha^i  \cdot (\mathbf{k}^i+\gamma^5 \mathbf{A}_j^i),
\end{eqnarray}
where now $\Alpha^i=(\openone \otimes \sigma^x, (-1)^i \sigma^z \otimes \sigma^y)$, $\gamma^5=\sigma^z \otimes \openone$ and $\mathbf{A}_1^i = \delta J_1 \frac{(a_i-1)}{a_i+1} (1, 1)$. This Hamiltonian describes the Dirac fermions being coupled to a chiral gauge field \cite{Jackiw}. As $\mathbf{k}^i=(0,0)$ no longer gives a vanishing Hamiltonian, the coupling to $\mathbf{A}^i_1$ shifts the paired Fermi points $\mathbf{Q}_+^i$ and $\mathbf{Q}_-^i$ towards each other, the direction being the same for both pairs. This agrees with $\delta H_{1}$ respecting both symmetries \rf{SLS1} and \rf{SLS2}, which also implies that all Fermi points have to vanish simultaneously. This is indeed the case as shown in Figure \ref{FP}(a), where we numerically demonstrate that dimerization in the large $\delta J_1$ limit can cause localization of the fermions on the $z$-links and thus completely remove the Fermi points. 

We can similarly study the transition from the $\nu=-2$ phase to the non-Abelian Ising phase ($\nu=-1$). This occurs when one tunes the couplings between the staggered, $(J_{z,1}, J_{z,2})=(-1,1)$, and the uniform, $(J_{z,1}, J_{z,2})=(1,1)$, configurations. In \rf{H1} this is equivalent to the Hamiltonian perturbation 
\be \label{dH2}
	\delta H_{2} = \delta J_2 \sigma^y \otimes (\sigma^x - \openone),
\ee
which respects the sublattice symmetry \rf{SLS1}, but breaks \rf{SLS2}, the emergent symmetry responsible for the chiral Abelian phase. The low-energy theory is again a Dirac field coupled to a chiral gauge field \rf{H4}, but now $\mathbf{A}_2^i = \frac{\delta J_2}{a_i+1} (1, (-1)^{i+1}\sqrt{3})$, which shifts the pairs $\mathbf{Q}^1_\pm$ and $\mathbf{Q}^2_\pm$ independent of each other. This is confirmed by Figure \ref{FP}(b), which shows that large $\delta J_2$ distortions can cause the $\mathbf{Q}_\pm^1$ Fermi points to annihilate while only transporting the other two. 

\begin{figure}[t, floatfix]
 \begin{tabular}{cc}
 \includegraphics*[width=6cm]{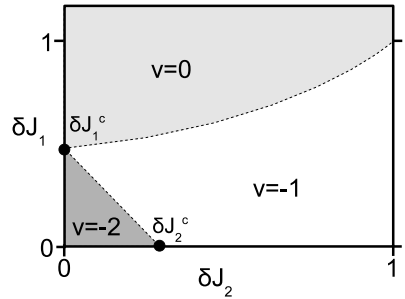} 
\end{tabular}
\caption{\label{phases} A section of the phase space as a function of $\delta J_1$ and $\delta J_2$. The dashed lines are the phase boundaries and the circles are the locations of the $K$ dependent critical points $\delta J_1^c = \sqrt{2+K^2}-1$ and $\delta J_2^c = \frac{1+K^2}{4}$. The phases characterized by the Chern number $\nu=0$, $\nu=-1$ and $\nu=-2$ correspond to the dimerized, uniform and staggered coupling configurations, respectively.}
\end{figure}

The study of the Fermi point transport holds also for $K \neq 0$, which allows us to obtain the $K$ dependent critical points $\delta J_i^c$. As $\delta J_i$ is varied, the gapped Fermi points (the minima/maxima of the bands $\pm E_{1,\mathbf{p}}$) follow the same trajectories, although slower for larger $K$. The annihilations still occur at the critical momenta $\mathbf{Q}_c=\{(0,0), (\frac{\pi}{2},0),(\frac{\pi}{2},\pi) \}$, where the gap always closes. This means that the eigenvalues of $H_{\mathbf{Q}_c}+\delta H_j$ must satisfy $\pm E_{1}(K,\delta J_j)=0$. The solutions are given by $1+\delta J_1^c = \sqrt{2+K^2}$ and $\delta J_2^c = \frac{1+K^2}{4}$, which agree with our previous numerical studies \cite{Lahtinen}. In Figure \ref{phases} we outline the extended phase space as functions of $\delta J_1$ and $\delta J_2$ showing the three distinct topological phases. For general values of the couplings, the critical staggered configuration and the tri-critical point occur when 
\be
J_{z,1}=\frac{K^2-1}{2}J_{z,2},
\ee
and
\be  
J_z^2 = J_x^2+J_y^2+K^2,
\ee
respectively. These extend the previous results \cite{Pachos06} and show that a larger $K$ has a stabilizing effect on the $\nu=-2$ phase.

\section{The role of anyon-anyon interactions}

In general, a transition from a non-Abelian Ising phase to a chiral Abelian phase has been predicted to arise due to anyon-anyon interactions \cite{Trebst, Read00-2}. As the Ising anyons in the honeycomb lattice model are strongly interacting \cite{Lahtinen}, we can explicitly establish this connection at an intuitive level.

\begin{figure}[t]
 \begin{tabular}{cccc}
  \includegraphics*[width=2cm]{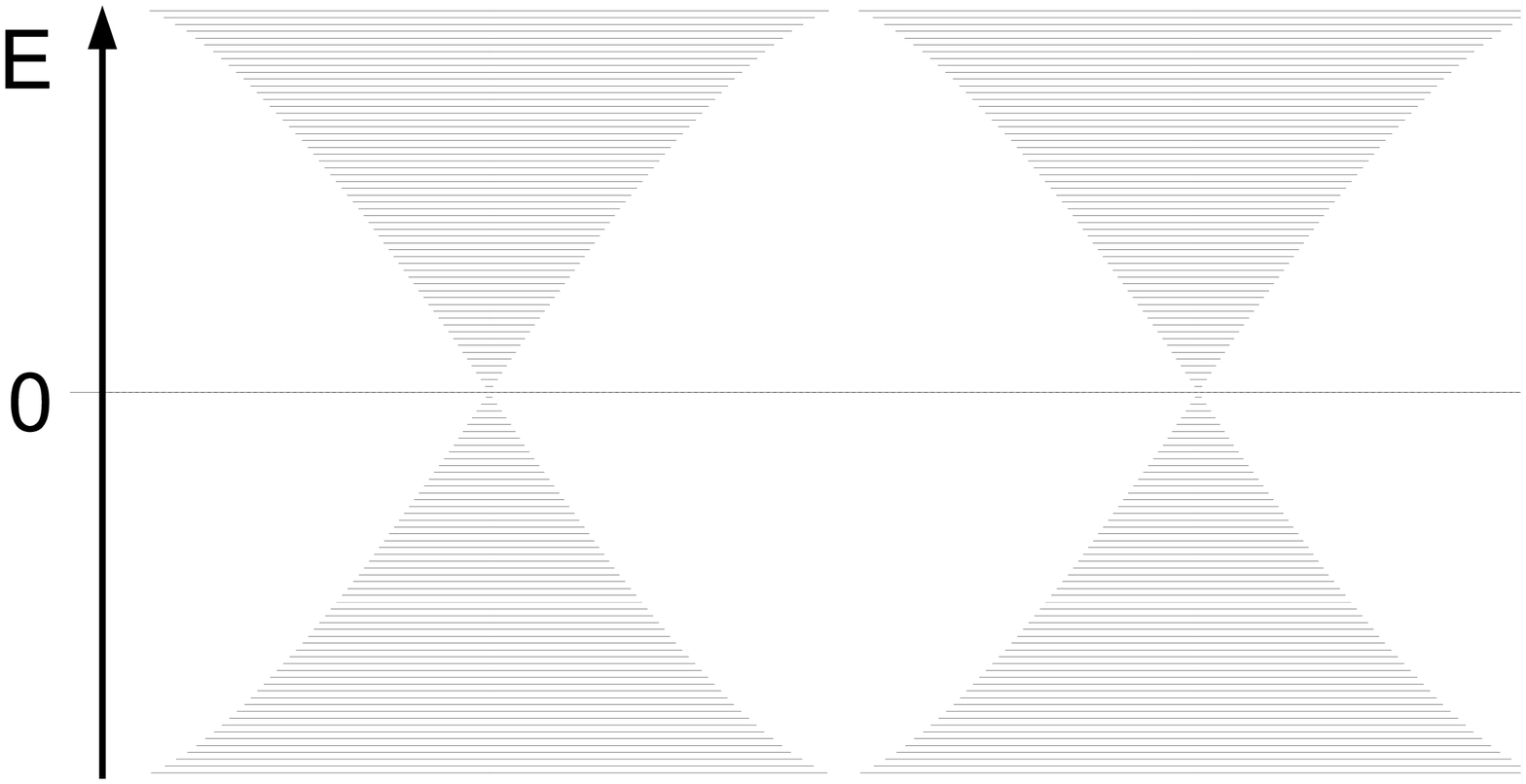} & \includegraphics*[width=2cm]{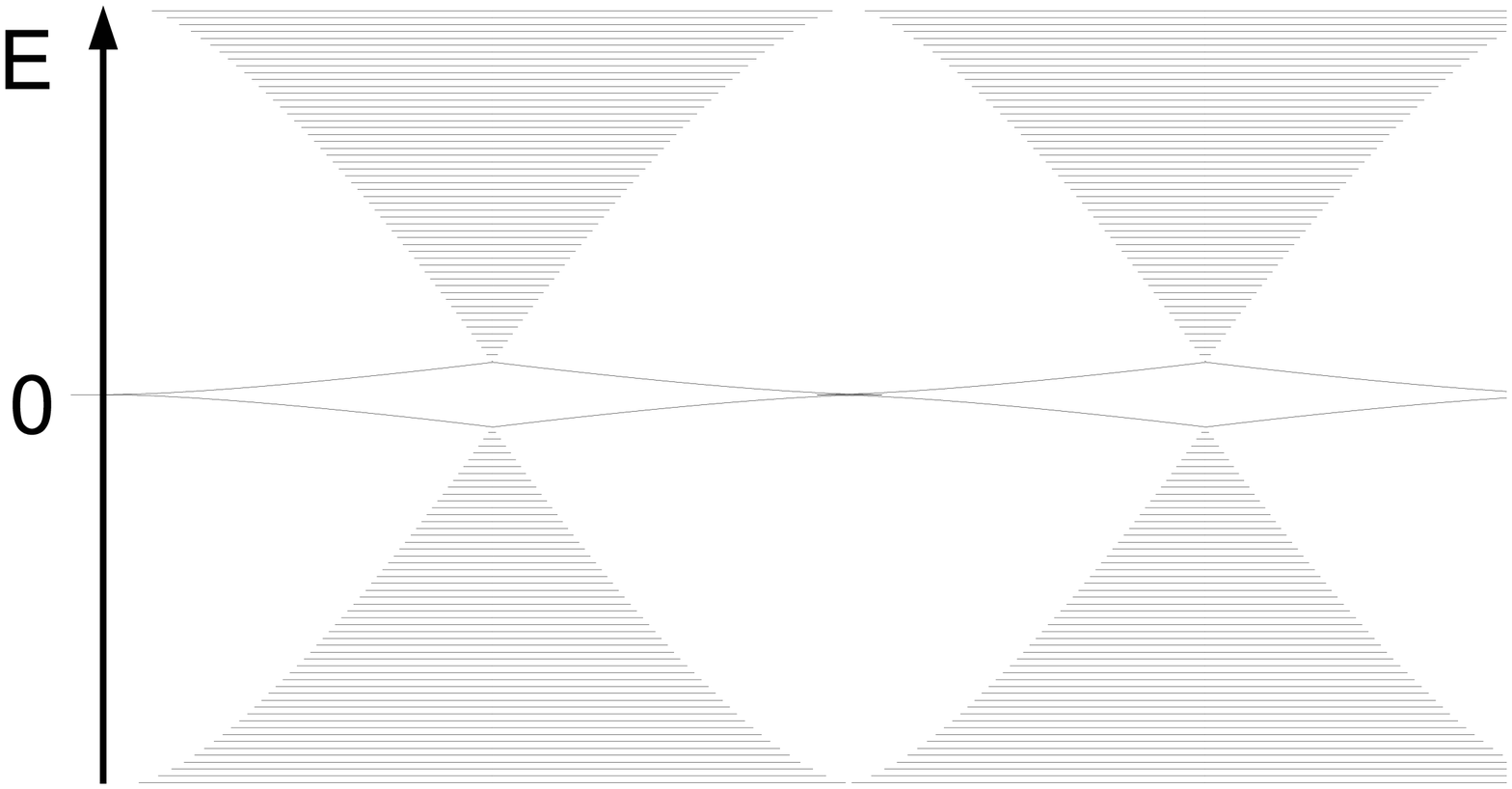} & \includegraphics*[width=2cm]{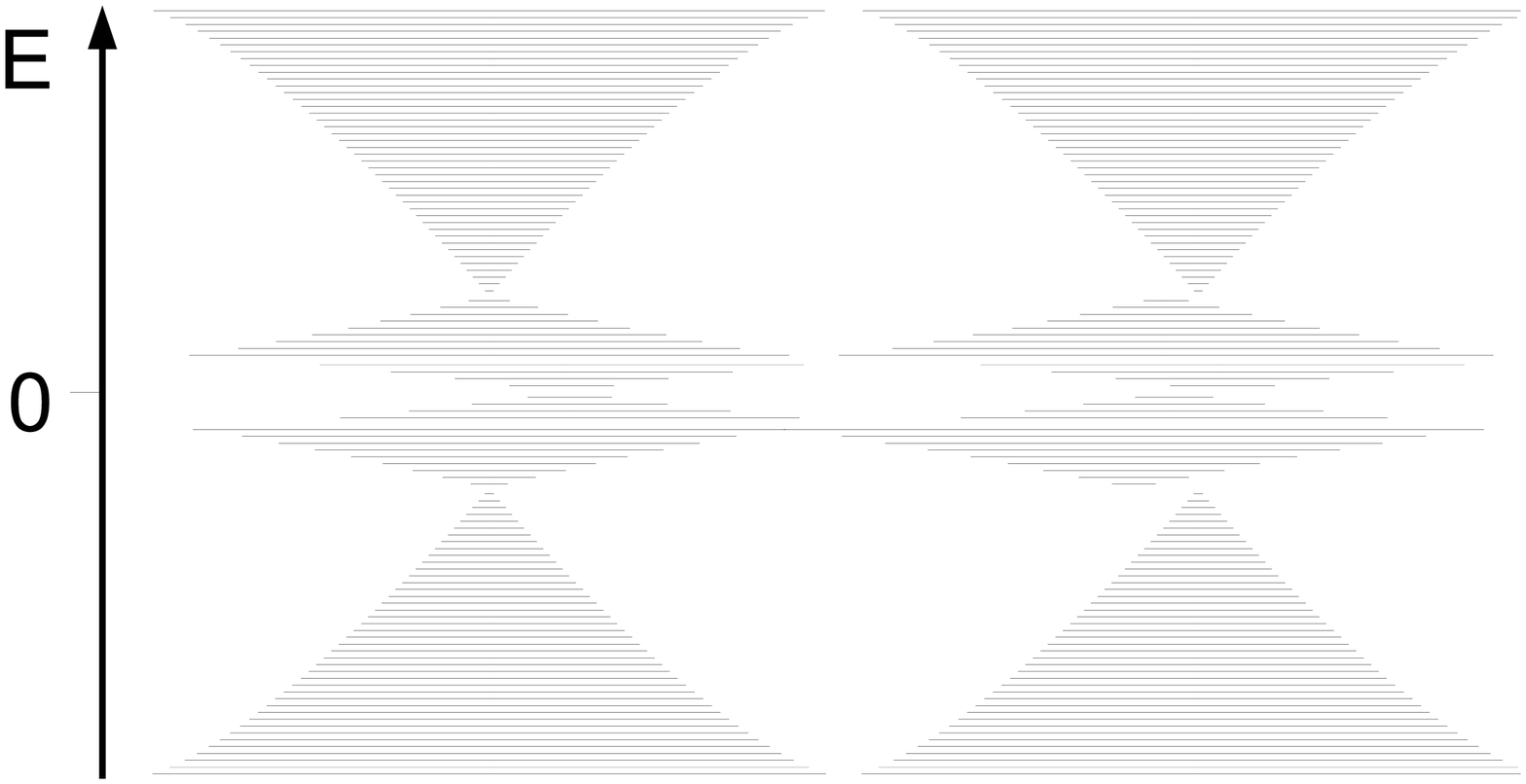} & \includegraphics*[width=2cm]{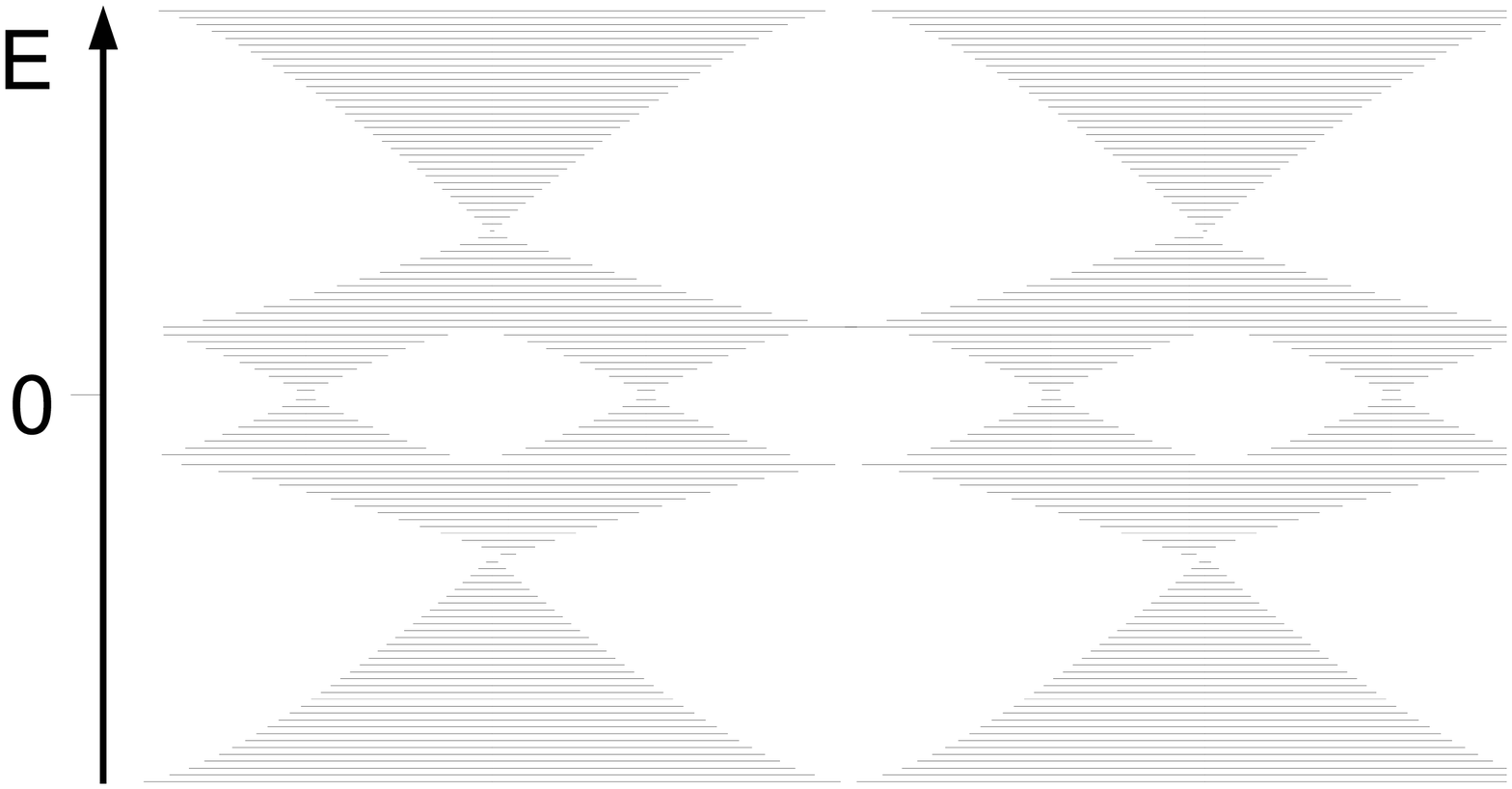} \\
 \includegraphics*[width=2cm]{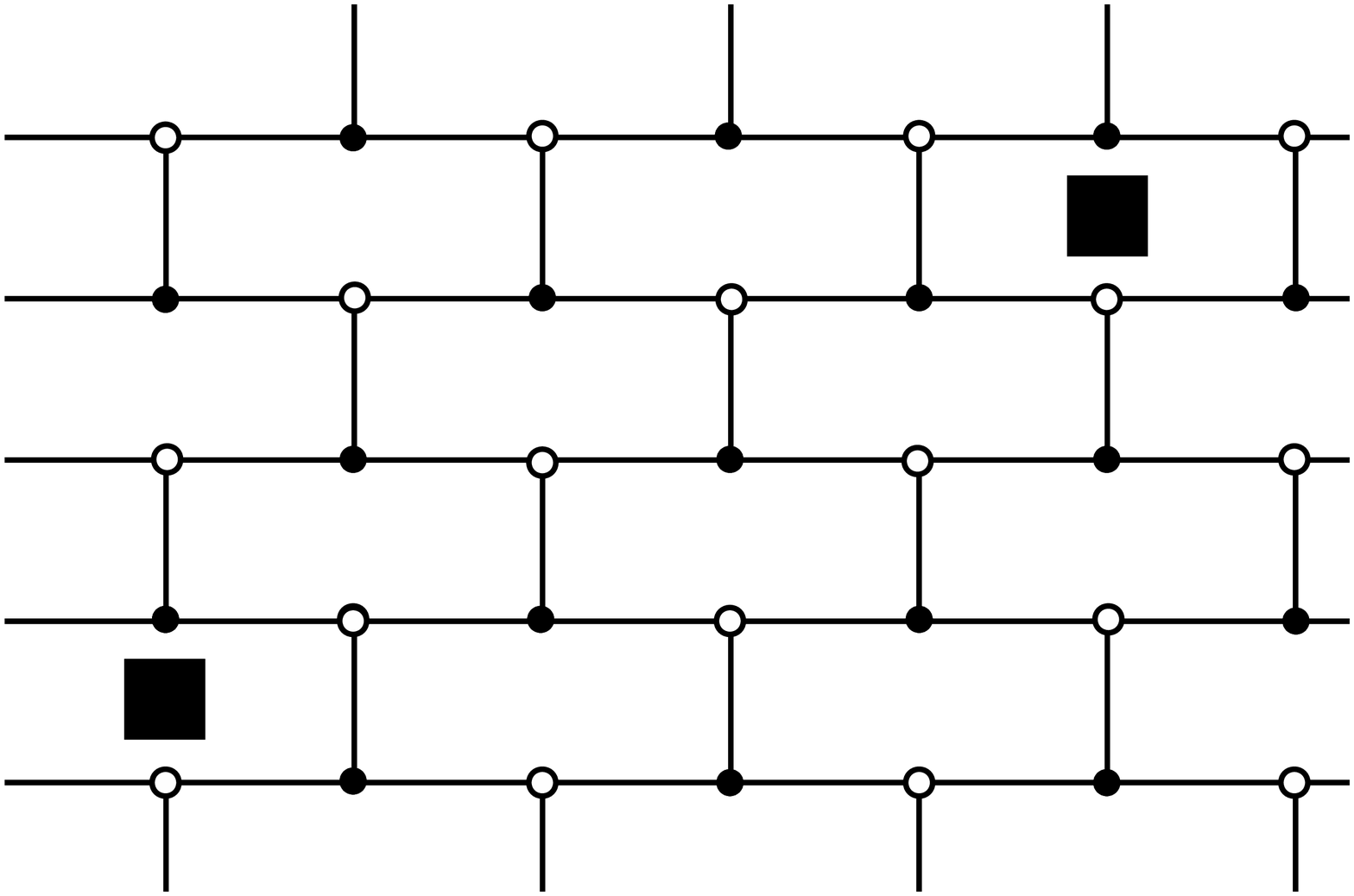} & \includegraphics*[width=2cm]{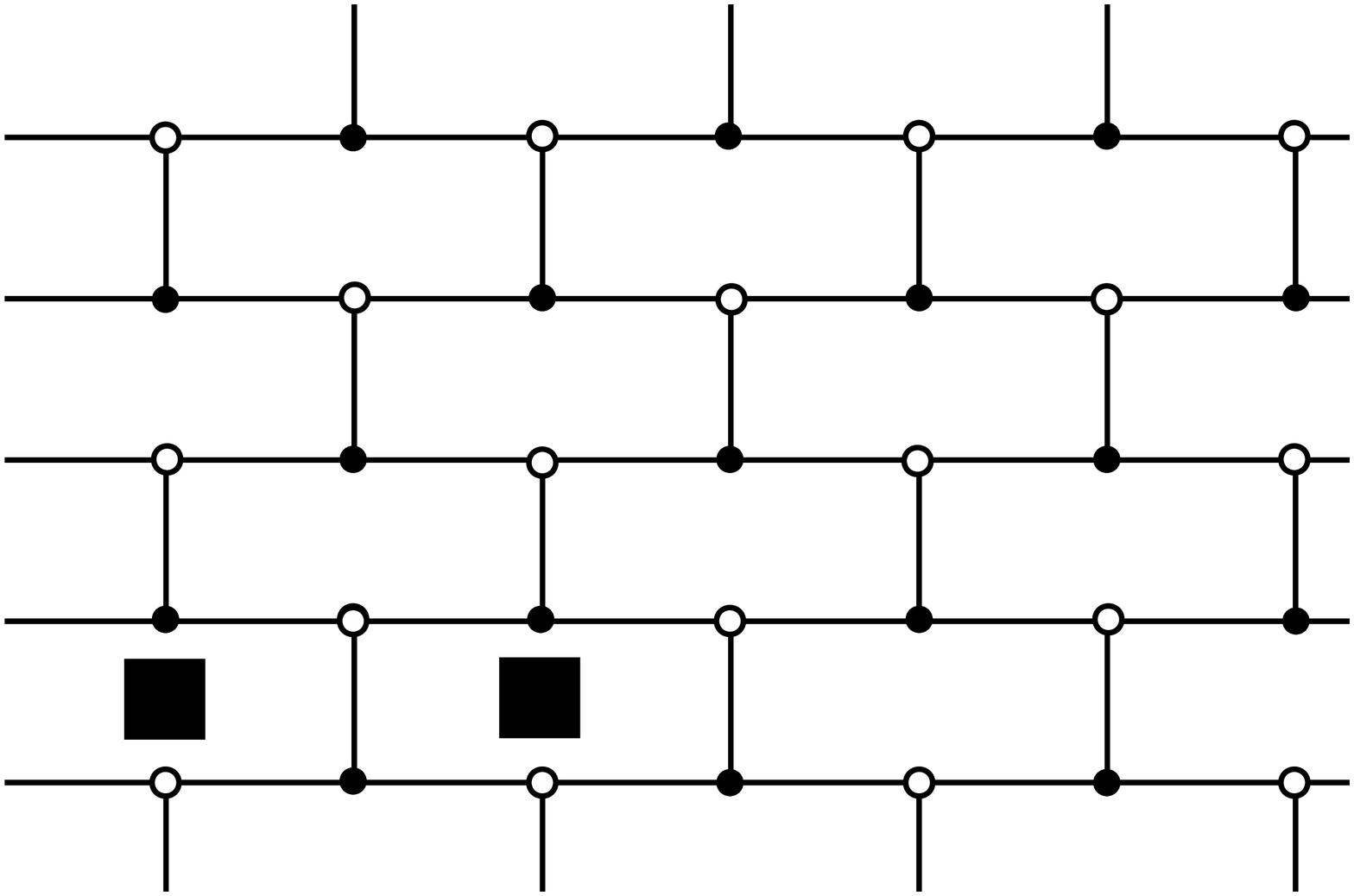} & \includegraphics*[width=2cm]{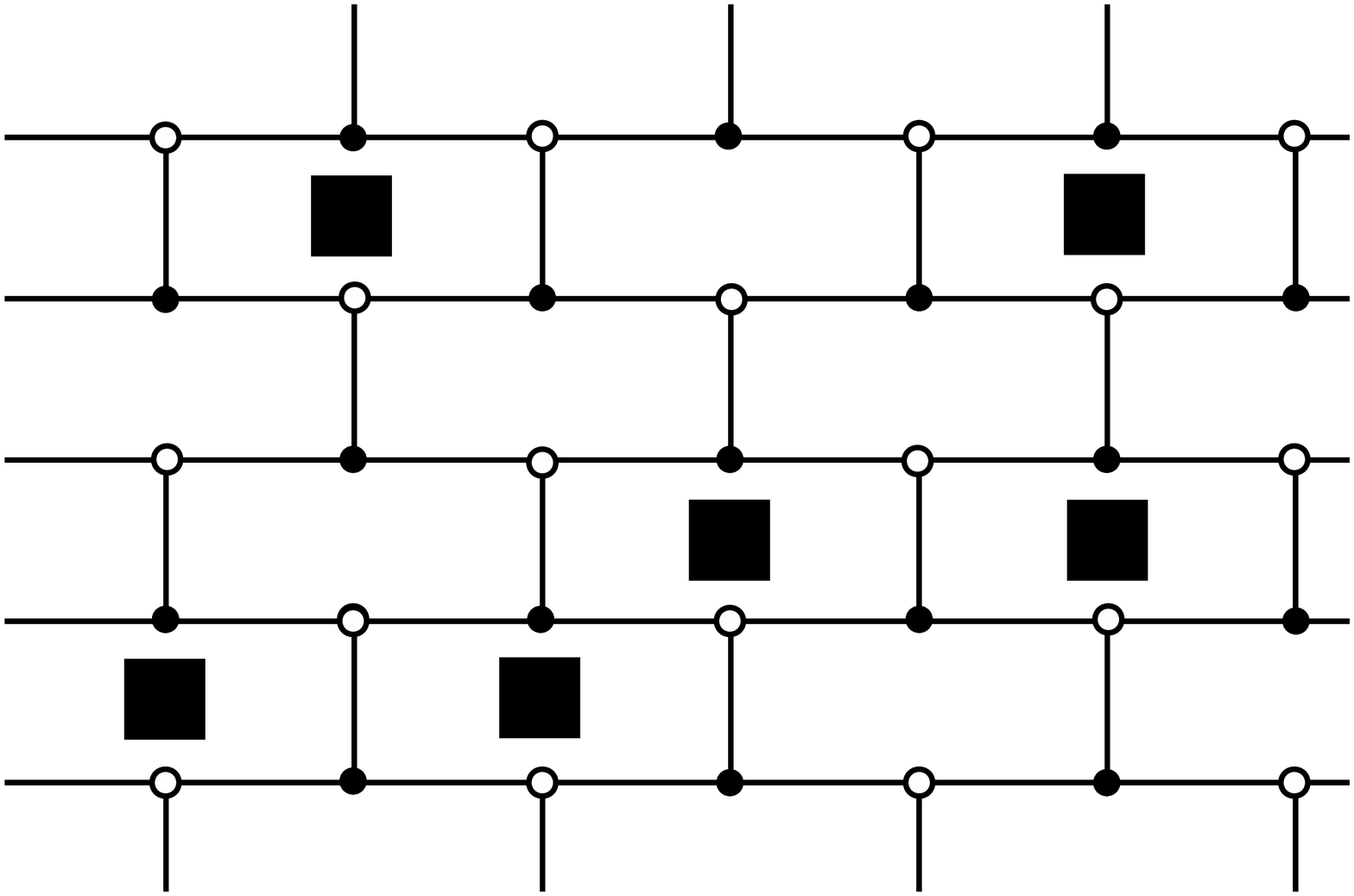} & \includegraphics*[width=2cm]{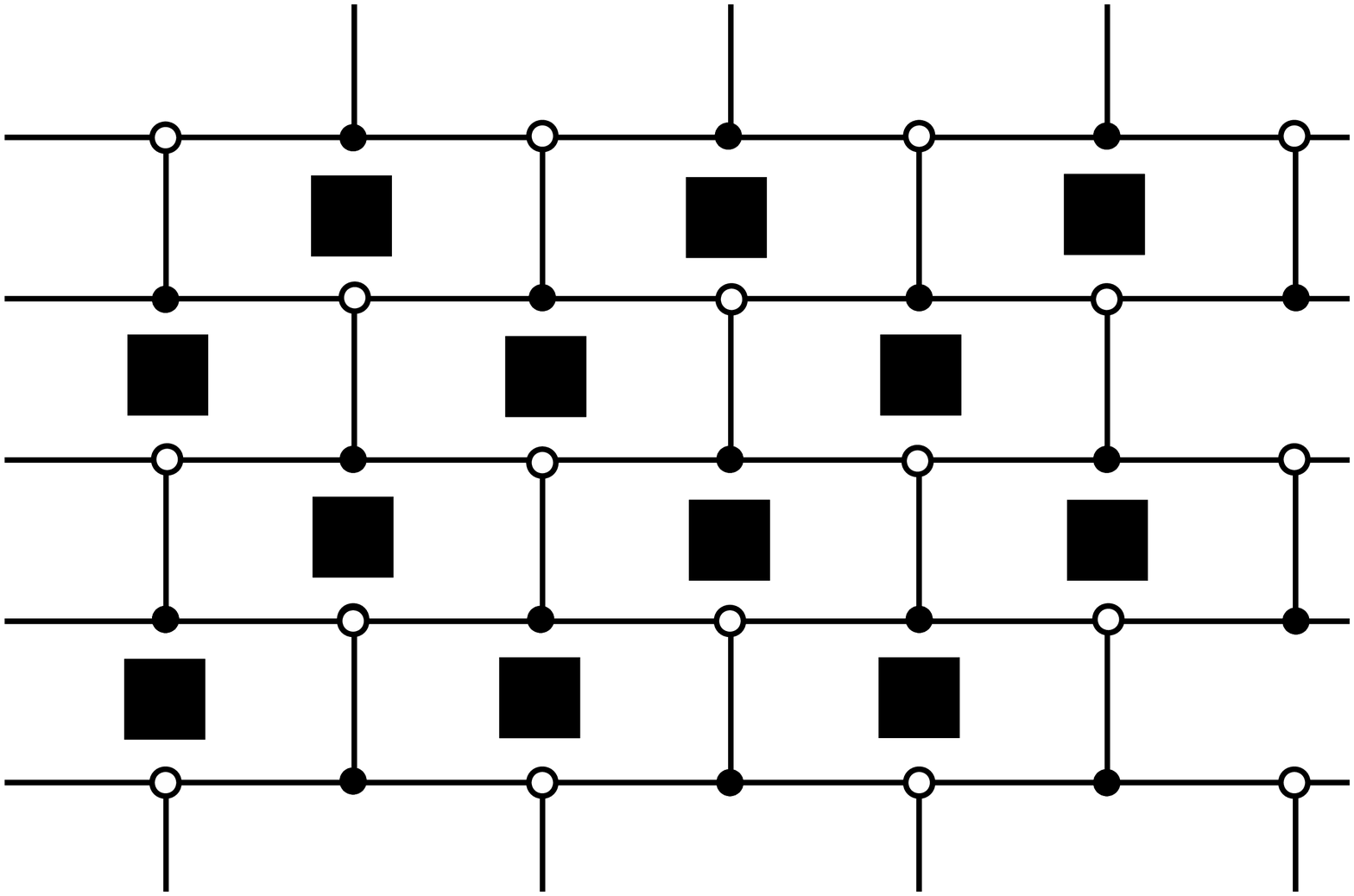} \\
(a) & (b) & (c) & (d)
\end{tabular}
\caption{\label{cartoon} A schematic illustration of the emergence of the full-vortex band structure due to interacting anyonic vortices (black squares) as the vortex density is increased. (a) A separated pair of vortices carries a zero mode ($E_{1,\mathbf{p}}=0$ for all $\mathbf{p}$). (b) Short-range interaction causes the zero mode to acquire momentum dependence. (c) The presence of many interacting vortices causes the zero modes to lose their character and to start forming a band structure. (d) The full-vortex band structure. }
\end{figure}

Based on numerical studies, we illustrate in Figures \ref{cartoon}(a)-(d) how the spectrum evolves as the vortex density inside a unit cell of fixed size is increased. Isolated vortices introduce modes with zero energy (Figure \ref{cartoon}(a)), which acquire momentum dependance at close ranges due to interactions (Figure \ref{cartoon}(b)). These \emph{zero modes} describe the anyonic fusion degrees of freedom \cite{Lahtinen}, that are unique to non-Abelian anyons. When many vortices interact with each other simultaneously, these modes start forming a band structure (Figure \ref{cartoon}(c)). Finally, as the density approaches the limiting full-vortex sector, the characteristic four Fermi points emerge (Figure \ref{cartoon}(d)).

We observe that the Fermi surface is modified due to an emergence of a new low-energy band that is purely due to the interacting non-Abelian anyons. As demonstrated earlier, it is directly related to the Chern number that characterizes the ground state of the whole system. Therefore, our study reveals the microscopic mechanism for the anyon-anyon interaction driven phase transitions. 

\section{Conclusions}

In summary, we have demonstrated the existence of a new topologically ordered phase in the honeycomb lattice model, which supports chiral Abelian anyons. This phase appears when the signs of the couplings alternate periodically, which is equivalent to creating a fully packed lattice of strongly interacting vortices. We characterized the phase by its distinct Fermi surface that arises due to an emergent lattice symmetry, and derived analytically the phase boundaries by studying the Fermi point transport. We showed analytically that the transport, and thus the topological phase transitions, translate in the low-energy theory to coupling to chiral gauge fields. Their form is shown to be directly related to the lattice symmetries, which could be employed when engineering gauge fields in laboratory \cite{Goldman09}, or studying the effects of lattice deformations in graphene \cite{Pereira}. 

The chiral Abelian phase can also be realized in the proposed optical lattice experiments, where the required staggered coupling configurations can be created by using several parallel lattices with readily tunable spin interaction patterns \cite{Micheli05}. Various experimental techniques can be considered for detecting the change in the Chern number \cite{Bermudez}. Furthermore, as the $\nu=-1$ phase with interacting non-Abelian anyons is equivalent to a $p$-wave superconductor \cite{Chen08}, a similar transition occurs also there for a sufficiently dense vortex lattice. Our results should apply as well to the chiral spin liquid variants of the model \cite{Yao07}.

\subsection*{Acknowledgements}

We would like to thank Simon Trebst and Andreas W. W. Ludwig for illuminating discussions. This work is supported by EPSRC, the Finnish Academy of Science, the EU Network EMALI and the Royal Society.

\end{document}